\documentclass[prb,twocolumn] {revtex4-1} 
\usepackage{graphicx}
\usepackage{url}
\usepackage{upgreek}
\begin{document}
\title{First measurements of the flux integral  with the NIST-4 watt balance}
\author{D.~Haddad}
\author{F.~Seifert}
\author{L.S.~Chao}
\author{A.~Cao}
\author{G.~Sineriz}
\author{J.R.~Pratt}
\author{D.B. Newell} 
\author{S.~Schlamminger}

\affiliation{Fundamental Electrical Measurements Group, National Institute of Standards and Technology, Gaithersburg, MD 20899}


\date{\today}



\begin{abstract}
In early 2014, construction of a new watt balance, named NIST-4, has started at the National Institute of Standards and Technology (NIST). In a watt balance, the gravitational force of an unknown mass is compensated by an electromagnetic force produced by a coil in a magnet system. The electromagnetic force depends on the current in the coil and the magnetic flux integral. Most watt balances feature an additional calibration mode, referred to as velocity mode, which allows one to measure the magnetic flux integral to high precision. In this article we describe first measurements of the flux integral in the new watt balance. We introduce measurement and data analysis techniques to assess the quality of the measurements and the adverse effects of vibrations on the instrument.
\end{abstract}
\maketitle
\section{Introduction}
Watt balances~\cite{kibble75} have a long history at the National Institute of Standards and Technology (NIST). In 2011, it was decided to build a new watt balance (NIST-4)~\cite{dh14} which  will be used to realize the kilogram once the international community has agreed to proceed with the redefinition of the international system of units, the Syst\`eme International d'Unit\'es (SI)~\cite{si,dn14}. In one aspect, the new watt balance is a departure from previous instruments~\cite{olsen80,steiner05b,ss14} built at NIST because a permanent magnet system instead of an electromagnet is used to generate the required magnetic field. This design choice is also used in other watt balances~\cite{gournay05,ms07,baumann13,sanchez14,kim14,sutton14} and leads to an instrument that is much smaller than its predecessors. 

In the watt balance experiment, the gravitational force, $F=mg$, on a mass, $m$, at the local gravitational acceleration, $g$, is counteracted by an electromagnetic force produced by a current carrying coil immersed in a magnetic field. The force generated by this coil can be expressed, in its simplest form, as $F=I lB$, where $I$ denotes the current in the coil, $B$ the average value of the magnetic flux density perpendicular to the wire and $l$ the wire length of the coil. 

Balancing the gravitational with an electromagnetic force is one of two measurements, sometimes called modes, that are performed with a watt balance. The purpose of the second mode is to precisely measure the flux integral $lB$. This measurement is implemented by moving the coil perpendicular through the magnetic field while recording the coil's velocity, $v$, and the induced voltage, $V$, in the coil. The ratio gives the flux integral, $lB=V/v$. 

\begin{figure}[t!]
\centering
\includegraphics[width=3.0in]{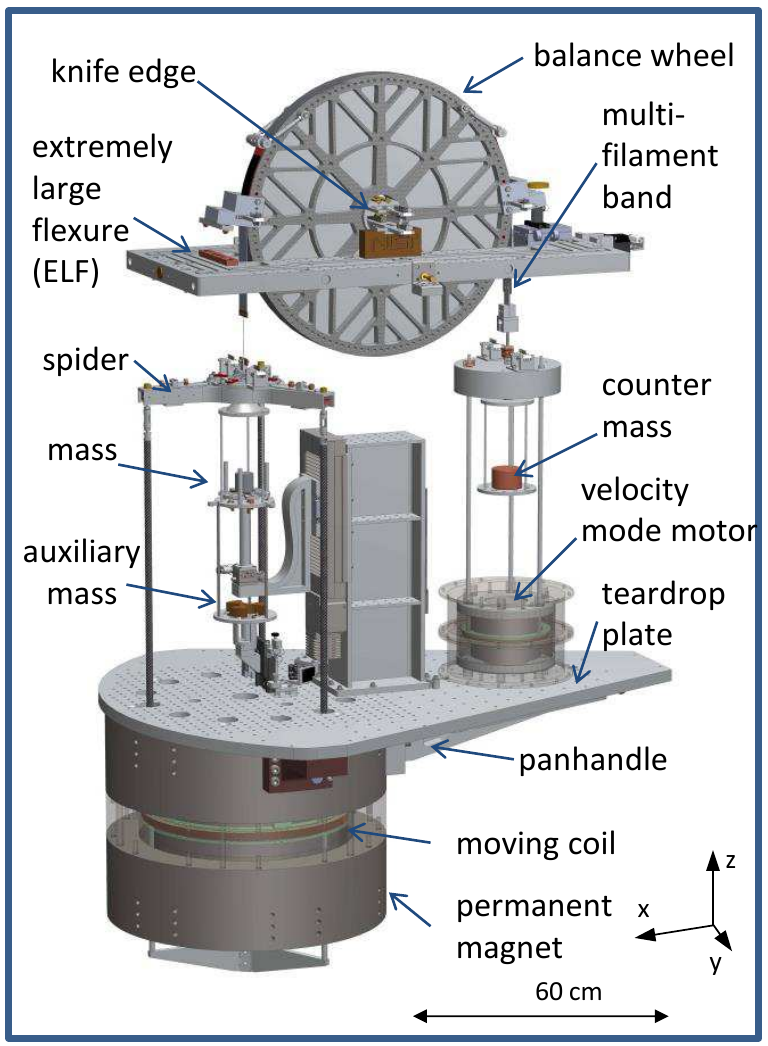}
\caption{Three dimensional drawing of the NIST-4 watt balance. For clarity the vacuum vessel and elements of the support structure have been omitted.}
\label{fig:3d_app}
\end{figure}

Construction of NIST-4 started after three years of design and preparation in January of 2014. Immediately after the balance was assembled, measurements of the flux integral were  taken with the goal to demonstrate that a measurement of the flux integral with a relative uncertainty of $10^{-8}$ is possible. The main concern for this measurement was the effect of mechanical vibrations of the instrument on the data. NIST-4 is smaller and lighter than NIST-3~\cite{ss14} leading to higher frequencies of its mechanical resonances. Preliminary measurement of the ground vibration at the NIST-4 site indicated larger amplitudes than at the NIST-3 site. In particular, the amplitude of the vertical ground motion at 29\,Hz at the NIST-4 site is about 30 times larger than at the NIST-3 site, see Fig.~\ref{fig:spectrum}. This vibration is most likely caused by electrical motors in large air conditioning units nearby. 

\section{The apparatus}

Figure~\ref{fig:3d_app} shows a drawing of NIST-4. Identical to the previous three instruments, a wheel serves as the balancing mechanism. In the current setup, an aluminum wheel with a diameter of 60.96\,cm and a thickness of 2.54\,cm is used. A polished stainless steel ring is press-fitted on the outside diameter of the aluminum wheel to provide a hard, polished surface for a multi-filament band to roll on. The multi-filament band is made from 65 Platinum-Tungsten wires.  The wheel is supported at its center from a 7.87\,cm long tungsten-carbide knife edge that can pivot  on a polished tungsten-carbide block. The center of mass of the wheel is adjusted to be closely beneath the knife edge. The free period of oscillation about the knife edge is about 10\,s.

The two sides of the wheel are referred to as the main-mass and the counter-mass side. On the main-mass side, the multi-filament band connects to a rotational coupler which is a bundle of ~150 straight Platinum-Tungsten wires each with a diameter of $76\,\mu$m and an overall length of 89\,mm.  The coupler terminates at a flexure system comprised of two independent flexures that pivot about the same point. One flexure supports the mass pan, while the other flexure supports the spider, a three-pointed star made from aluminum. The spider connects to the coil by three carbon fiber rods that are mounted with two flexures on either end allowing the coil to move parallel to the spider. 

When the wheel is rotating, the multi-filament band rolls on or off the stainless steel surface shortening or lengthening the free hanging part of the band. Even though the band was annealed, internal stresses produce torques about the vertical axis on the lower end of the band. If not counteracted, these torques would produce an unwanted rotational oscillation of the coil about the $z$-axis. 
An electrostatic torque compensation system is used to mitigate this problem. The actuator consists of  six high voltage electrodes, three for clockwise and three for counter clockwise torques, surrounding three grounded gold plated glass plates mounted on the carbon fiber rods. The input of the PID loop of the torque compensation system is calculated from two optical sensors, each monitoring the $x$ and $y$ position of a retro-reflector mounted on the coil former. From these data, the rotation of the coil around the $z$ axis can be calculated. Such a system was already used successfully in the NIST-3 experiment. When the PID loop is off, the amplitude of the rotation can be adjusted with the rotational coupler to within 0.1\,mrad when the coil moves up and down. With the PID loop on, the amplitude is better than 0.01\,mrad. The rotational coupler is an element with low torsional stiffness that allows us to servo the rotational orientation of the coil with moderate voltages (below 1000\,V).

\begin{figure}[t!]
\centering
\includegraphics[width=3in]{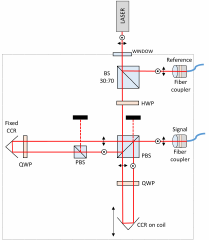}
\caption{Topology of the heterodyne interferometer used to measure the vertical position and velocity of the coil. The dotted box denotes the vacuum envelope. Legend: BS =  beam-splitter, PBS = polarizing beam splitter, HWP = half-wave plate,  QWP = quarter wave plate, CCR = corner cube retro-reflector. }
\label{fig:ifo}
\end{figure}

The coil former  is made from glass-epoxy laminate, grade FR-4, the material from which printed circuit boards are typically made. The coil former contains two rectangular grooves, vertically displaced by 11.5\,mm. In each groove, a coil with 464 turns and a mean radius of 21.75\,cm is wound, using AWG-24 wire (diameter approximately 0.5\,mm), yielding a resistance of about 56.2\,$\Omega$. Electrically both coils are connected in series.

The coil is inside a permanent magnet system, where two Samarium-Cobalt, Sm$_2$Co$_{17}$, disks source the magnetic field and soft iron (AISI 1021) guides the flux through a 3\,cm wide annular gap. The magnetic flux density in the center of the 15\,cm high gap is 0.55\,T. The mass of the magnet system is approximately 816\,kg. More details on the design and construction of the permanent magnet system can be found in references~\cite{ss13} and \cite{fs14}.

From the coil, three aluminum posts extend down to a second spider which supports a corner cube reflector in its center. This  reflector is used to interferometrically measure the coil's vertical position and velocity. The light beam of the interferometer is traveling through a central bore in the magnet.

On the counter-mass side of the wheel, a smaller coil is suspended in a magnet system to provide a small drive force for the velocity control. This coil has a nominal diameter of 18.3\,cm 297 turns, with a resistance of 21\,$\Omega$. The coil is inside a smaller magnet system that uses Neodymium-Iron-Boron, Nd$_2$Fe$_{14}$B, permanent magnets to produce a flux of 0.14\,T in the center of a 3\,cm wide gap. In the middle of the 10.16\,cm high gap, the flux integral is 23.6\,T\,m.  This counter-mass coil is connected by three rods to an aluminum cylinder that hangs on a multi-filament band off the wheel. The aluminum cylinder is necessary to counter-balance the heavier main mass assembly.

The vertical position of the coil is measured with a heterodyne interferometer using an Agilent 5517C laser\footnote{Certain commercial equipment, instruments, or materials are identified in this paper in order to specify the experimental procedure adequately. Such identification is not intended to imply recommendation or endorsement by the National Institute of Standards and Technology, nor is it intended to imply that the materials or equipment identified are necessarily the best available for the purpose.} as the light source. The optical layout, shown in Figure~\ref{fig:ifo}, was designed to minimize polarization mixing~\cite{bobroff87}. The half-wave plate in the optical path before the polarizing beam splitter allows us to align the polarization directions of the light beam  to the polarization axes of the polarizing beam splitter cube.  The additional polarizing beam splitter in the reference arm reduces the amount of horizontally polarized light leaking in the reference arm of the interferometer, reducing the error produced by polarization mixing. The effect of this additional beam splitter can be clearly observed in the amplitude spectral density of the measured velocity and  reduces the amplitude spectral density by about a factor of two.

\begin{figure}[t!]
\centering
\includegraphics[width=3in]{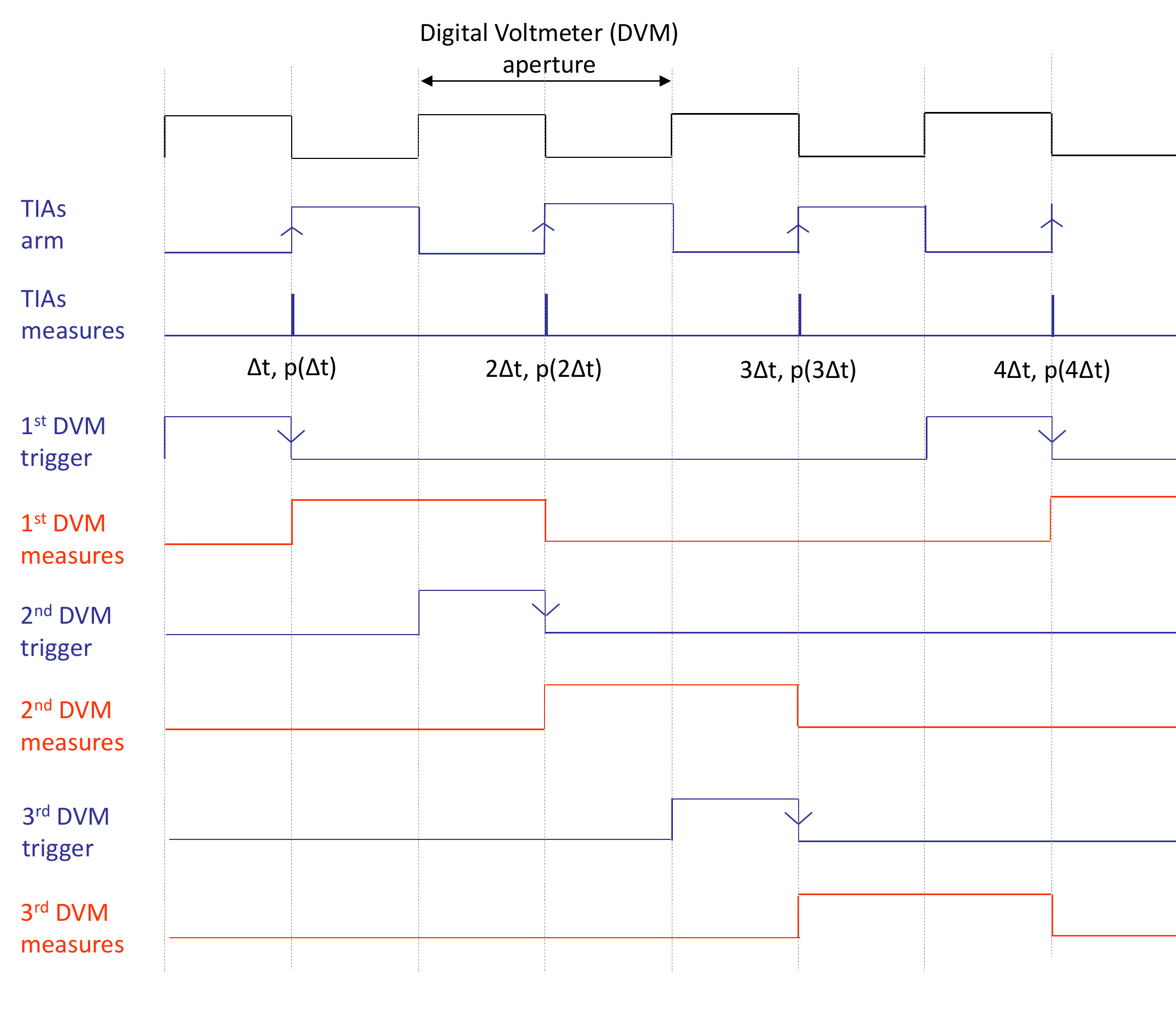}
\caption{Timing diagram for the time interval analyzer (TIA) and the three digital voltmeters used in the experiment.}
\label{fig:timing}
\end{figure}

The signals from two optical pickups, one for the reference and one for the interferometer signal, are converted to an electrical signal with a photo receiver (Agilent E1709A). The electrical signals are fed into a time interval analyzer (TIA) made by Brilliant Instruments, Model BI221. The time interval analyzer is triggered by the electrical pulses received from the reference. The signal to arm the trigger is generated by a pulse generator using a frequency up to 300\,Hz, see Fig.~\ref{fig:timing}. The triggers for three digital voltmeters (DVMs) that are used to sample the induced voltage in the coil are generated at the same frequency, but a time delay is adjusted to obtain simultaneous measurements of voltage and velocity. This delay was adjusted to within 500\,ns by moving the coil in a sinusoidal waveform and analyzing the Lissajous figures obtained by plotting voltage versus velocity. 

\begin{figure*}[t!]
\centering
\includegraphics[width=6in]{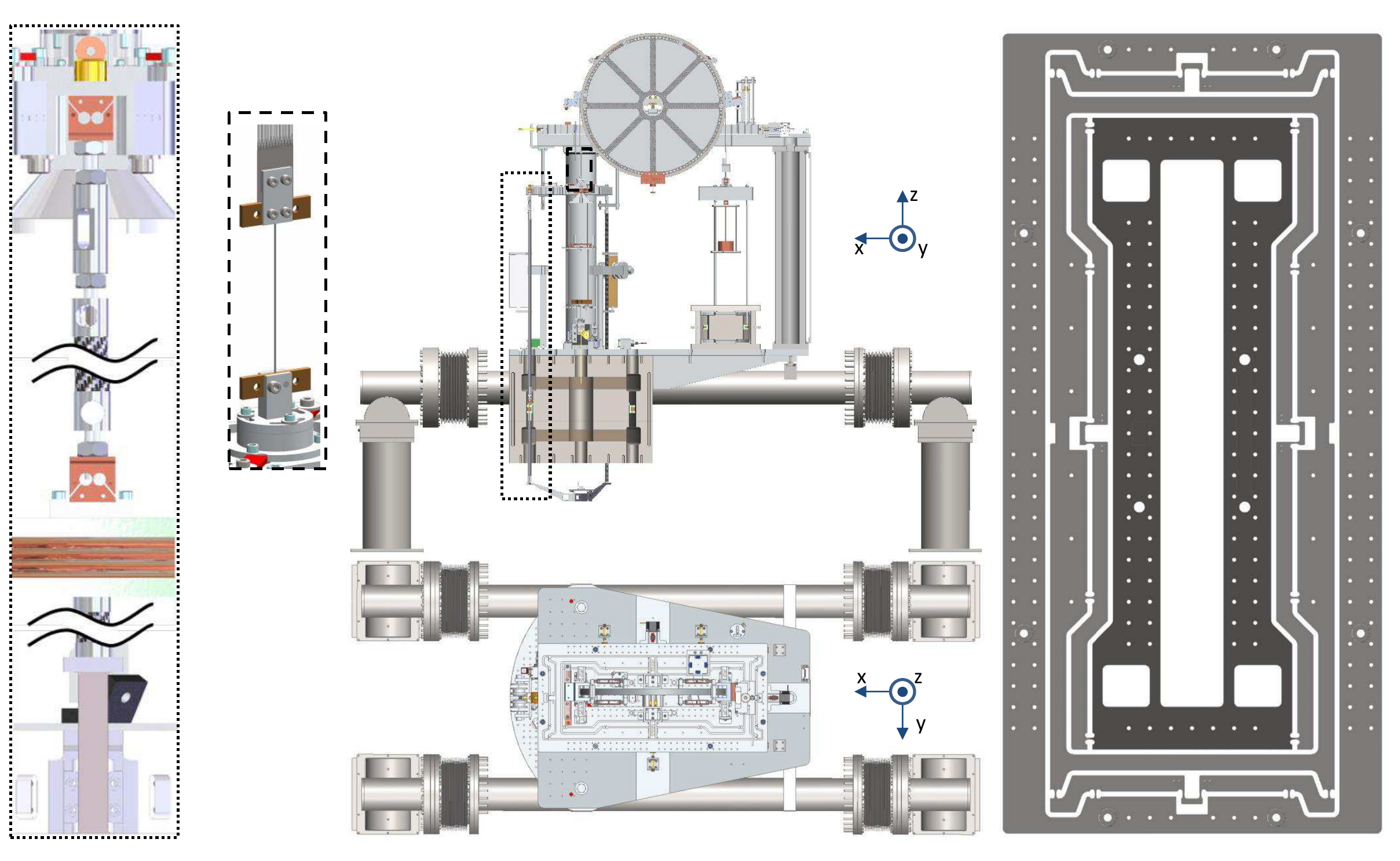}
\caption{The support system of the watt balance. In the middle, top: A cross-sectional drawing as seen from the side. In the middle, bottom: A top view of the instrument. For simplicity the vacuum vessel is omitted. Other than through the four bellows shown, there is no connection between the balance and the vacuum vessel. The directions of our coordinate system are indicated in the graph. On the left in the dotted box: Enlarged truncated drawing of one of the three coil support rods. On the left in the dashed box: The termination of the multi-filament band and rotational coupler. On the right: A top view of the extremely large flexure (ELF). The outer ring is bolted to the trapezoidal plate. The inner part of the ELF, drawn with darker color, can move in x- and y- direction by $\pm$0.5\,mm. }
\label{fig:susp}
\end{figure*}

To understand the vibrational motion of the coil with respect to the magnet system, a closer look at the suspension of the watt balance is necessary. A cross-sectional side view and top view of the main suspension elements can be found in Figure~\ref{fig:susp}. The main watt balance rests on two sand filled steel tubes  (inner diameter 17.8\,cm, outer diameter 20.3\,cm) which are supported by four legs. These legs are located on the corners of a rectangle with a length of 2.33\,m and width of 0.81\,m. The magnet engages in a  pseudo-kinematic mount onto the beams. Two support points are mounted at the top diameter of the magnet, the third on a panhandle that rests on a bridge between the two beams. A large aluminum plate in the shape of a teardrop is bolted to the top surface of the magnet. This aluminum plate is the base for mounting the optics and is the building platform for the mass lift (not shown in Fig.~\ref{fig:susp}) . Three posts, again sand filled, stand on this plate and support the table which is a trapezoidal shaped plate with a central slot to accommodate the wheel. On top of the table, rests the extremely large flexure (ELF). The ELF allows us to move the knife edge support  by $\pm$0.5\,mm along the x- and y-direction. These degrees of freedom are necessary to align the coil concentric with the magnetic field in order to reduce the torques on the coil during the weighing mode. A detailed description of the ELF can be found in~\cite{lc13}.

The main vibrational frequencies in the balance listed in order of decreasing frequencies are:
\begin{itemize}
\item 30\,Hz. The suspension of the coil on the main mass side, consisting of the bands, the spider, and the carbon fiber rods has a bounce frequency of 30\,Hz.
\item 21\,Hz and its harmonics. The moving part (drawn in dark color in Fig.~\ref{fig:susp}) of the ELF is bouncing at this frequency. To damp this vibration the inner part has been pretensioned and damped with bolts. While this damps the 21\,Hz vibration, it introduces vibrations at harmonics, particularly at 63\,Hz. In the future, we plan to insert damping material between the table and the ELF.
\item  16\,Hz. The watt balance on the two steel beams moves  up and down at a frequency of 16\,Hz. 
\item 1.7\,Hz. This frequency is the double pendulum motion from the point where the band rolls off from the wheel. In this mode, the spider moves in one direction and the coil in the other.
\item 0.85\,Hz and 0.834\,Hz. The tilt of the coil oscillates about its center of mass at two frequencies around the x- and y-axis, respectively. These motions are referred to as wobble motions of the coil.
\item 0.56\,Hz. The pendulum motion of the coil, more precisely the shearing motion between the coil and the upper spider. For this motion, all six flexures on the carbon fiber rods are bending.
\end{itemize}
The frequencies listed above were measured with the wheel at the middle position in its travel range. Rotating the wheel will slightly change some of these frequencies due to shortening or lengthening of the suspension.

\section{Data analysis and first results}

The induced voltage, $V$,  in the coil is proportional to its vertical velocity, $v$, and the flux integral, $lB$. Because of the finite size of the magnet, the magnetic flux varies along the gap and thus the flux integral is a function of coil position. Hence, $V(t,z) = v(t) lB (z)$.  Neglecting the non-uniformity of the flux integral, during the watt experiment, the coil moves with a nominal velocity  $v_{\mathrm{nom}}$ and a nominal induced voltage $V_{\mathrm{nom}}  \approx v_{\mathrm{nom}}  lB$ is measured. Here, $lB$ is 697\,T\,m and thus for a nominal induced voltage of 1\,V a velocity of 1.43\,mm/s is required.

For a stationary balance, i.e., a non-rotating wheel, the residual coil movement is a small vibration around an average position $z_o$. In this case, the excursions from the nominal positions are small and the position dependence of $lB$ can be neglected. However, the quotient of voltage to velocity cannot be calculated because the velocity will sometimes be zero driving the quotient to infinity. Instead,  we introduce a quantity, named the residual,
\begin{equation}
\varepsilon(t) = V(t)-v(t) lB(z_o).
\end{equation}
For a vibrationally driven coil and sufficiently long observation time, the expectation value of $\varepsilon(t)$ is $\langle \varepsilon(t)\rangle = 0$. The $lB(z_o)$ can be calculated from the slope of the measured voltage as a function of velocity, $V(v)$. The residual is useful in two ways. First, in the frequency domain, it allows one to identify the regions where the motions of the coil track the induced voltage and where they don't.  Second, as we will show below, from measurements of $\varepsilon$ an estimate of the uncertainty of the $lB$ measurement in the real experiment can be calculated as follows.

In velocity mode of a watt balance experiment, one tries to estimate the flux integral by measuring the induced voltage and the velocity of the coil, i.e.
\begin{equation}
lB = \frac{V}{v}.
\end{equation}
 The relative uncertainty squared of the determination of the flux integral  in the watt balance experiment can be written as 
\begin{equation}
\frac{\sigma_{lB}^2}{(lB)^2}  = \frac{\sigma_V^2}{V_{\mathrm{nom}}^2} +\frac{\sigma_v^2}{v_{\mathrm{nom}}^2} - 2 r_{vV} \frac{\sigma_V}{V_{\mathrm{nom}}}\frac{\sigma_v}{v_{\mathrm{nom}}} 
\end{equation}
where $r_{vV}$ denotes the correlation coefficient between the voltage measurement and the velocity measurement. The squared value of the absolute uncertainty of $\varepsilon$ is given by
\begin{equation}
\sigma_{\varepsilon}^2 = \sigma_V^2 +(lB)^2\frac{\sigma_v^2}{v^2} - 2 lB r_{vV} \sigma_V \sigma_v. \label{eq:eps}
\end{equation}
Dividing this equation by the squared value of nominally induced voltage during the watt experiment, $V_{\mathrm{nom}}^2$ and using the relationship, $V_{\mathrm{nom}} = lB  v_{\mathrm{nom}}$, yields
\begin{equation}
\frac{\sigma_{\varepsilon}^2}{V_{\mathrm{nom}}^2} = \frac{\sigma_V^2}{V_{\mathrm{nom}}^2} +\frac{\sigma_v^2}{v_{\mathrm{nom}}^2} - 2 r_{vV} \frac{\sigma_V}{V_{\mathrm{nom}}} \frac{\sigma_v}{v_{\mathrm{nom}}} = \frac{\sigma_{lB}^2}{(lB)^2} .
\end{equation}
The relative uncertainty achieved for measuring $lB$ is the same as the quotient of the absolute uncertainty of $\varepsilon$ divided by the nominal induced voltage.

\begin{figure}[htb]
\centering
\includegraphics[width=3.3in]{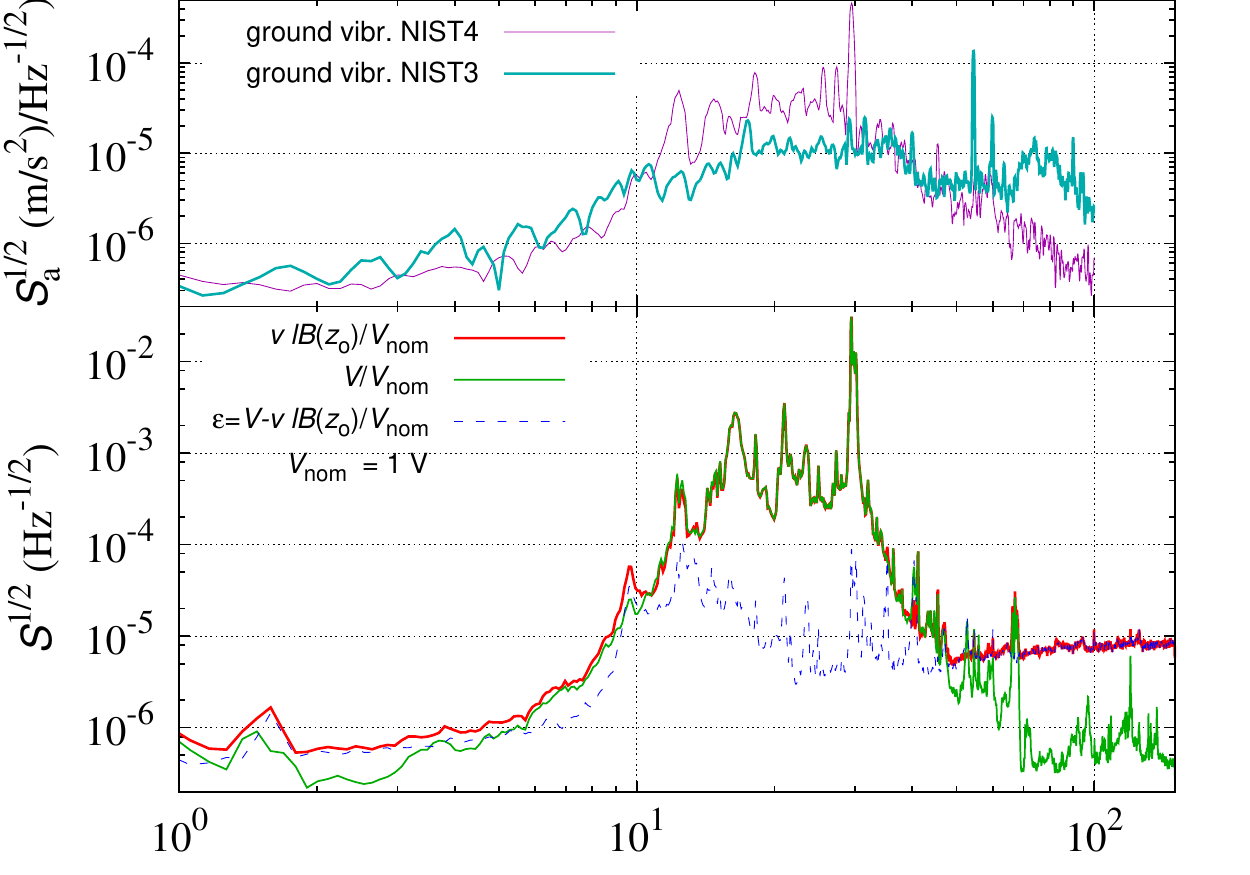}
\caption{Top Panel: Amplitude spectral density of the ground vibrations measured at the site of the new watt balance (NIST-4) and the old watt balance (NIST-3). Bottom Panel: Amplitude spectral density of the induced voltage $V$, the velocity multiplied with the flux integral, $v lB$, and the residual $\varepsilon$. All three spectra are divided by $V_{\mathrm{nom}}=1$\,V. }
\label{fig:spectrum}
\end{figure}

The amplitude spectral density of the three quantities, $V$, $vlB$, and $\varepsilon$ are shown in figure~\ref{fig:spectrum}. All  three spectral amplitudes are divided by $V_{\mathrm{nom}}$=1\,V.  These three spectra correspond to the amplitude spectral density of the square roots of the term on left and the first two terms on the right of the equal sign in Eq.~\ref{eq:eps}. All three spectra share the same vertical unit, $1/\sqrt{\mbox{Hz}}$. The traces for the spectra of $V/V_{\mathrm{nom}}$ and $vlB/V_{\mathrm{nom}}$ are on top of each other for a large range in frequency. At these frequencies, both signals are in common mode and the amplitude spectral density of $\varepsilon/V_{\mathrm{nom}}$ is much smaller than either one. With increasing frequency at about 8\,Hz, the noise in the velocity measurement exceeds the noise in the voltage measurement and dominates the amplitude spectral density, i.e., the amplitude spectral density of the residual follows that of the velocity. In general, the amplitude spectral density of the velocity increases with increasing frequency, because the velocity is the derivative (multiply by $f$ in frequency domain) of the position measurement, which exhibits a nearly flat amplitude spectral density.

\begin{figure}[htb]
\centering
\includegraphics[width=3.3in]{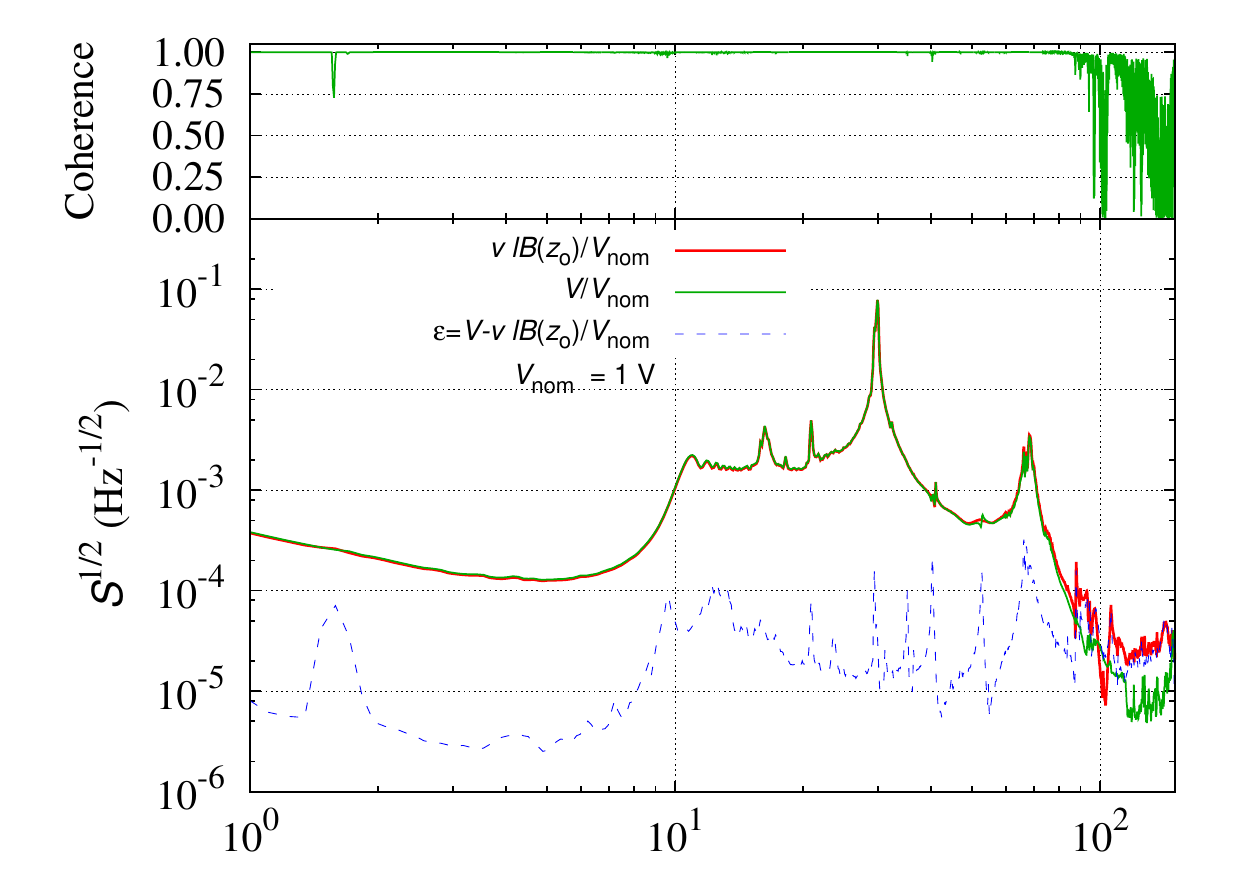}
\caption{Lower graph: The amplitude spectral density of the induced voltage, the coil's velocity and the residual while the counter mass coil is driven by a swept sine wave with a frequency ranging from 1\,Hz to 150\,Hz and a 0.5 V amplitude. The normalization of the spectra is the same as in Fig.~\ref{fig:spectrum} Upper graph: coherence between voltage and velocity during this measurement.}
\label{fig:RespCoh}
\end{figure}

In order to investigate the basic resonances of the watt balance, a sinusoidal voltage with an amplitude of 0.5\,V and a frequency sweeping from 1\,Hz to 150\,Hz was applied to the counter mass coil, while the induced voltage in the main coil and the velocity of the main coil was recorded. Fig.~\ref{fig:RespCoh} shows the amplitude spectral densities of the voltage, the velocity, and the residuals with the same normalization as above. Two strong resonances can be seen, one at 30\,Hz, the other at 67\, Hz. Both resonances are double peaks. The 30\, Hz is the up down vibrational mode of the coil caused by stretching of the support. The  voltage and velocity are in phase and coherent up to 80\,Hz. At 80\,Hz, the coherence~\cite{bendat} decreases, see upper graph in Fig.~\ref{fig:RespCoh}.

When the balance is servoed to a fixed position or moved with constant velocity, the twin peaks at 30\,Hz  dominate the amplitude spectral density of the residual and hence the amplitude spectral density of the flux integral. These signals make it difficult to obtain a precise measurement of $lB$. Using longer integration times on the voltmeter and correspondingly larger sample times for the time interval analyzer did not solve the problem. In this case, the peaks are aliased into the measurement band introducing spurious signals. A promising solution is to sample the DVMs and the TIAs with a relatively large frequency (120 Hz) and to apply a digital, finite impulse response  (FIR) low-pass filter to both channels before calculating the residuals or the ratio. Currently, we use a filter with a passband-edge frequency of about 6\,Hz and an attenuation of at least 110\,dB for frequencies larger than a stopband-edge frequency of 10\,Hz.

\begin{figure}[htb]
\centering
\includegraphics[width=3.3in]{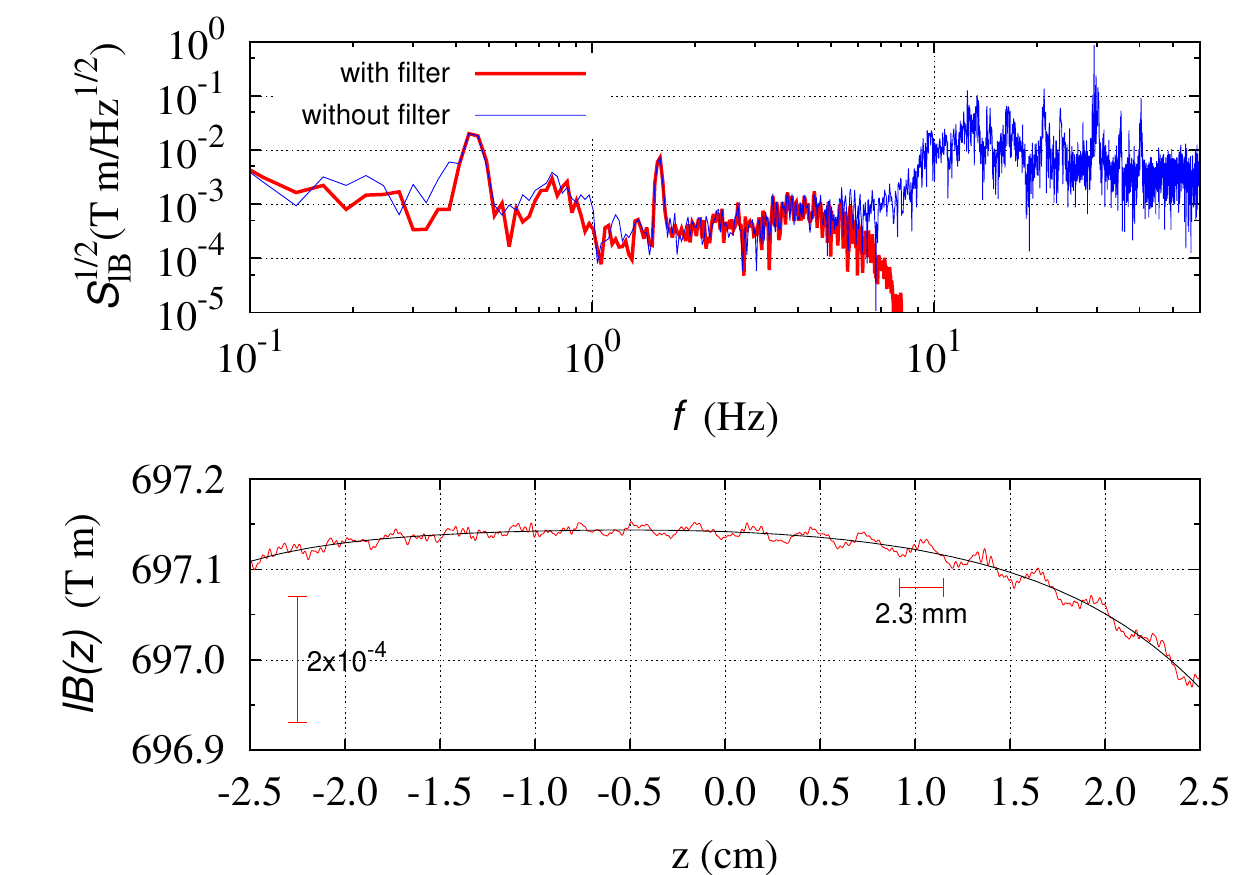}
\caption{ Top graph: The amplitude spectral density of the measured $lB$ during a constant velocity motion of the coil.  The $lB$ is calculated from the ratio of the measured voltage and velocity. In one case, dashed line, the ratio is calculated from the as-measured data stream. In the second case, solid line, the voltage and velocity data is passed through a digital FIR filter. The 3 dB corner frequency of the filter is at about 6\,Hz. Bottom graph: The flux integral calculated from the filtered data from one 5\,cm long motion through the magnetic field.}
\label{fig:profile}
\end{figure}

To measure the flux integral, the coil is moved through the magnet. The balance is in servo control  using the velocity mode motor as actuator. The set-point of the position servo varies with a triangular waveform with an amplitude of 3\,cm and a period of 96\,s. To avoid bias at the turning points, the data analysis is limited to the inner 5\,cm of travel, where the coil moves with a constant velocity of 1.36\,mm/s. Figure~\ref{fig:profile} shows the data of one such sweep. The $lB$ is calculated by dividing the induced voltage by the measured velocity. In the top panel, the amplitude spectral density of the $lB$ is shown for the filtered and unfiltered data stream. The amplitude in the unfiltered data is dominated by the amplitude density beneath the  twin peaks at 29\,Hz and elevated white noise from 10\,Hz to the Nyquist frequency of 60\,Hz. 

After filtering the data, three dominant peaks remain at 0.55\,Hz, 0.85\,Hz and 1.7\,Hz, corresponding to the pendulum, wobble, and double-pendulum motions. These motions do not cancel in the ratio due to misalignment of the electrical center with the optical center.

 A common alignment practice is a two step procedure. In one step,  the  optical center  is aligned to the  mass center  by observing coil wobble and minimizing the second harmonic in the optical signal. The difference between optical center and mass center is referred to as Abbe offset In another step, the electrical center is aligned to the mass center by injecting current in the coil and observing the coil tilt. After the two step procedure, the electrical center is aligned with the optical center. In this preliminary setup, only a coarse alignment of the optical center with the mass center was possible.  In the future, we plan to use three interferometers to track three points on the coil former, each pair spaced 120$^o$ apart. By mathematically combining the readout of the three interferometers, the optical center is obtained. Changing the relative weighting of the three interferometer signals allows us to move the optical center in software, as required by one of the steps above. 

In addition to reducing the Abbe offset, two other strategies to reduce the effect of parasitic motions on the measurement result have been used in the past: In NIST-3, active damping is used before and after each sweep~\cite{steiner05a}. In the NRC balance, the starting time of the sweep is randomized with respect to the phase of the tilt motion~\cite{ian12}.

From the filtered data, the flux integral as a function of coil position is calculated and shown in the lower plot of Figure~\ref{fig:profile}. The measured variation of the flux integral as a function of position agrees well with measurements performed while characterizing the magnet as described in Ref.~\cite{fs14}. The  relative amplitude of the variation is $2\times 10^{-4}$.

\begin{figure}[htb]
\centering
\includegraphics[width=3.3in]{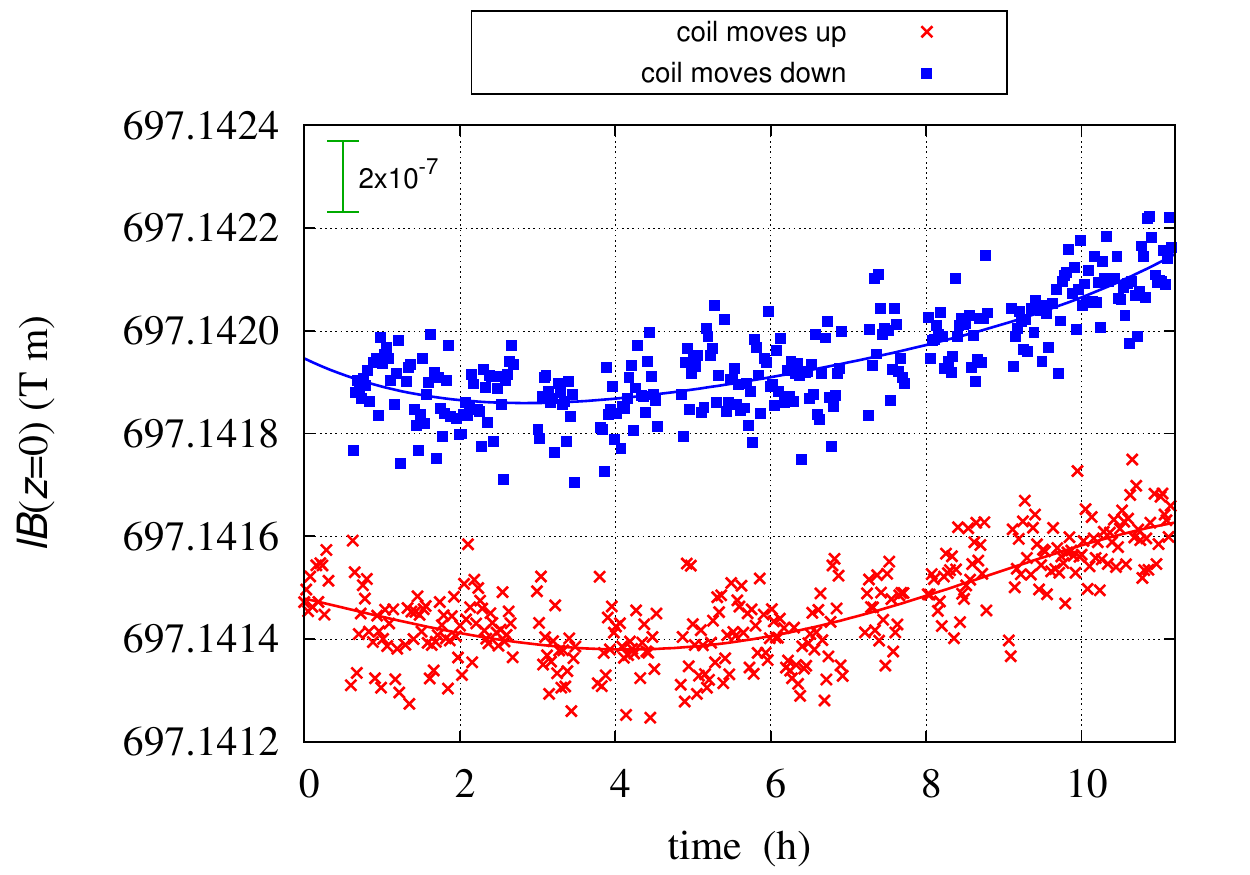}
\caption{Measurement of the flux integral throughout one night.}
\label{fig:bl}
\end{figure}

At low frequency, the amplitude spectral density of the flux integral is about $10^{-3}$\,T\,m/$\sqrt{\mbox{Hz}}$. Integrating this level over a 48\,s long sweep yields a relative uncertainty in $lB$ of $2\times 10^{-7}$. To verify this level of uncertainty, the coil was moved through the magnet for an entire night.
A sixth degree polynomial $lB(z)$ is adjusted in a least squares fashion to the data obtained in each sweep, see the black line in the lower plot of Fig.~\ref{fig:profile}. 
 This polynomial is then used to calculate $lB(0)$.  A plot of these numbers is shown in Figure~\ref{fig:bl}.  The results for the coil moving in one direction differ from that by moving in the other direction due to thermal voltages in the system. Ref.~\cite{ss14} shows the influence of thermal voltages in velocity mode. These thermal voltages, however, seem to remain fairly constant over the course of the night. Overall the flux integral drifts by a few parts in $10^{-7}$. This drift is caused by temperature changes of the Sm$_2$Co$_{17}$ in the permanent magnet. These data were taken several days after the vacuum system had been pumped down, when the magnet has reached a fairly stable temperature. The relative scatter in the data is close to the prediction  from the amplitude spectral density, about $1.3\times 10^{-7}$. Hence, a measurement of $lB$ with a relative statistical uncertainty of $1.3\times 10^{-8}$ should be possible in about 4,800\,s at a velocity of 1.36\,mm/s, assuming that the noise is white and stationary. At present, a relative statistical uncertainty of $1\times 10^{-8}$ would be achievable within 2.25 hours of measurement time.

\section{Conclusion}

The main components of the new NIST watt balance have been assembled. First measurements of the flux integral were performed and the noise levels and structural resonances of the system were thoroughly investigated.

We found that the coil motion has one strong vibrational resonance at 29\,Hz. In this mode the coil moves up and down, stretching the bands and carbon fiber rods. Upon close inspection, both resonance peaks split into two peaks.  These resonances dominate the ratio of volt to velocity, i.e., the flux integral. One successful way to remove excess noise from the data stream is to measure volt and velocity with a sample rate of 120\,Hz and then use a digital low pass filter with a cut-off frequency of 6\,Hz. From the two filtered data channels, the ratio is then calculated. Since both channels are filtered through the same digital filter, no bias is added to the data. This procedure will, however, add a phase delay  to the final values of $lB(t)$. 

In addition the performance of the system can be enhanced by improving the alignment. Especially by using three interferometers to allow a alignment of the Abbe offset.

With the current system, we have shown that it is possible to measure the flux integral with a relative statistical uncertainty of $1.3\times 10^{-7}$ for one sweep.

In the future, we will implement the force mode and incorporate a Josephson voltage standard in the system. We hope to have a fully working watt balance by summer 2015.


\begin{thebibliography}{1}

\bibitem{kibble75} Kibble B P, "A measurement of the gyromagnetic ratio of
the proton by the strong field method" Atomic Masses and Fundamental Constants vol. 5, ed J H Sanders and A H Wapstra (New York: Plenum), pp.~545-51, 1976.

\bibitem{dh14} D.~Haddad, L.S.~Chao, F.~Seifert, D.B.~Newell, J.R.~Pratt, S.~Schlamminger, "Construction of a watt balance with the aim to realize the kilogram at the National Institute of Standards and Technology", {\it Proc. CPEM. Dig.}, pp.~708--709, Aug.~2014.

\bibitem{si} International Bureau of Weights and Measures (BIPM), "The International System of Units (SI)", 8th edition (2006), http://www.bipm.org/en/si/si\_brochure.


\bibitem{dn14} D.B.~Newell, "A more fundamental International System of Units," {\it Physics Today}, vol.~87, no.~7 pp.~35--41, Jul.~2014.


\bibitem{olsen80} P.T.~Olsen, W.D.~Phillips and E.R.~Williams, "A proposed coil system for the improved realization of the absolute Ampere," {\it J. Res. NBS}, vol.~85, pp.~257--72, Jul.~1980.

\bibitem{steiner05b} R.L.~Steiner, E.R.~Williams, D.B.~Newell and R.~Liu, "Towards an electronic kilogram: an improved measurement of the Planck constant and electron mass," {\it Metrologia}, vol.~42, pp.~431--41, Sep.~2005.

\bibitem{ss14} S.~Schlamminger, D.~Haddad, F.~Seifert, L.S.~Chao, D.B.~Newell, R.~Liu, R.L.~Steiner and J.R.~Pratt, "Determination of the Planck constant using a watt balance with a superconducting magnet system at the National Institute of Standards and Technology," {\it Metrologia}, vol.~51, pp.~S15--S24, Mar.~2014.


\bibitem{gournay05} P.~Gournay, G.~Genev\`{e}s, F.~Alves, M.~Besbes, F.~Villar, and J.~David, "Magnetic Circuit Design for the BNM Watt Balance Experiment," {\it IEEE Trans. Instrum. Meas.}, vol.~54, no.~2, pp.~742--745, Apr.~2005.

\bibitem{ms07} M. Stock, "Watt balances and the future of the kilogram," {\it INFOSIM Informative Bulletin of the Inter American Metrology System}, vol.~9, pp.~9--13, Nov.~2006.

\bibitem{baumann13} H.~Baumann, A.~Eichenberger, F.~Cosandier, B.~Jeckelmann, R.~Clavel, D.~Reber and D.~Tommasini, "Design of the new METAS watt balance experiment Mark II," {\it Metrologia}, vol.~50, pp.~235--242, May~2013.

\bibitem{sanchez14} C.A.~Sanchez, B.M.~Wood, R.G.~Green, J.O.~Liard and D.~Inglis, "A determination of Planck’s constant using the NRC watt balance," {\it Metrologia}, vol.~51, pp.~S5--S14, Mar.~2014.

\bibitem{kim14} D.~Kim, B.-C.~Woo, K.-C.~Lee, K.-B.~Choi, J.-A.~Kim, J.W.~Kim and J.~Kim, "Design of the KRISS watt balance," {\it Metrologia}, vol.~51, pp.~S96--S100, Mar.~2014.

\bibitem{sutton14} C.M.~Sutton and M.T.~Clarkson, "A magnet system for the MSL watt balance," {\it Metrologia}, vol.~51, pp.~S101--S106, Mar.~2014.

\bibitem{ss13} S. Schlamminger, "Design of the Permanent-Magnet System for NIST-4," {\it IEEE Trans. Instrum. Meas.}, vol.~62, no.~6, pp.~1524--530, Jun.~2013.

\bibitem{fs14} F.~Seifert, A.~Panna, S.~Li, B.~Han, L.~Chao, A.~Cao, D.~Haddad, H.~Choi, L.~Haley, S.~Schlamminger, "Construction, Measurement, Shimming, and Performance of the NIST-4 Magnet System," {\it IEEE Trans. Instrum. Meas.}, vol.~63, no.~12, pp.~3027--3038, Dec.~2014.

\bibitem{lc13} L.S.~Chao, S.~Schlamminger, J.R.~Pratt, "Functional constraints and the design of a new watt balance," {\it Proc. ASPE An. Meetings}, pp.~209--213, Sep.~2013.

\bibitem{bendat} J.S. Bendat and A.G. Piersol, {\it Random Data: Analysis and Measurement Procedures}, 3rd ed. New York, NY, USA: John Wiley \& Sons, 2000.

\bibitem{bobroff87} N.~Bobroff, "Residual errors in laser interferometry from air turbulence and nonlinearity," {\it Appl. Opt.}, vol.~26, pp.~2676--2682, Jul.~1987.

\bibitem{steiner05a} R.~Steiner, D.~Newell, E.~Williams, "Details of the 1998 Watt Balance Experiment Determining the Planck Constant,"  {\it J. Res. Natl. Inst. Technol.}, vol.~110, pp.~1--26, Feb.~2005.

\bibitem{ian12} I.A.~Robinson, "Towards the redefinition of the kilogram: a measurement of the Planck constant using the NPL Mark II watt balance," {\it Metrologia}, vol.~49, pp.~113--156, Nov.~2011.

\end{thebibliography}
\end{document}